\documentclass[a4paper,twocolumn,english,aps,showpacs,letter]{revtex4}

\usepackage{amsmath}
\usepackage{graphics,color}
\begin{document}

\title[FES out of equilibrium]{Fermi edge singularity
in a non-equilibrium system}

\author{B. Muzykantskii$^1$, N. d'Ambrumenil$^{1,2}$ and 
B. Braunecker$^{3,1}$ }

\affiliation{$^1$Department of Physics, University of Warwick, Coventry CV4 7AL,
 UK \\
$^2$School of Physics and Astronomy, University of Birmingham, Birmingham B15 2TT, UK \\
$^3$Institute of Theoretical Physics, EPFL, CH-1015 Lausanne, Switzerland 
}

\date{\today{}}

\begin{abstract}
We report exact results for the
Fermi Edge Singularity in the  absorption 
spectrum  of an out-of-equilibrium tunnel junction. We consider
two metals with chemical potential difference $V$ separated by a
tunneling barrier containing
a defect, which exists in one of two states.
When it is in its excited state, tunneling
through the otherwise impermeable barrier is possible.
We find that the lineshape 
not only depends on the total scattering phase shift
as in the equilibrium case but also on the difference in the phase
of the reflection amplitudes on the two sides of the barrier.
The out-of-equilibrium spectrum extends below the original threshold
as energy can be provided by the power source driving current
across the barrier. 
Our results have a surprisingly simple interpretation
in terms of known results for the equilibrium case but with 
(in general complex-valued) combinations of elements of the 
scattering matrix  replacing the equilibrium phase shifts.
\end{abstract}
\maketitle

Developments in the fabrication and manipulation of
mesoscopic systems have allowed detailed and well-characterized
transport measurements for a large range of devices including
quantum pumps, tunnel junctions and carbon nanotubes.
It is often the case that such measurements explore non-equilibrium
effects particularly when the  potential
difference is dropped across a narrow potential barrier
or over a short distance inside the metallic region 
\cite{RB94,KGG01,NCL00}.
While there is often a very good theoretical description of much that
has been observed for systems close to equilibrium,
the theoretical picture for systems out of equilibrium 
is less clear with fewer established theoretical results.

A natural point to start, when looking for a description
of non-equilibrium effects in many-electron systems is
the Fermi Edge Singularity (FES), 
which is one of the simplest non-trivial many-body effects.
The FES is characteristic of the 
response of a Fermi gas to a rapid switching process. Initially
it was associated with the shape of
the absorption edge and
spectral line found when a core hole is created 
\cite{Mahan67}. However, it
turns out to be a generic feature of a Fermi system's response to
any fast switching process and reflects the large number of 
low-energy (particle-hole) excitations which exist in Fermi liquids.
It has also been shown to be
related to Anderson's orthogonality catastrophe 
\cite{ND69,CN71}
and can be used
to reformulate the Kondo problem in terms of a succession of spin 
flips which are treated as the switching of a one-body potential between
two different values \cite{YA70}.

\begin{figure}
{\centering     
\input{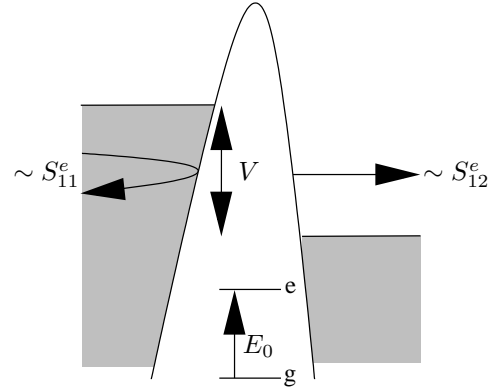} }

        \caption{Energy levels in an 
idealized device to demonstrate the out-of-equilibrium
FES. The scattering potential for electrons is characterized via the
$2\times2$  matrix,  $S(\epsilon,t)$,  connecting scattering states in the
two wires for particles with energy $\epsilon$.
$S=S^g$ or $S^e$ depending on whether
the defect is in its ground ($g$) or excited ($e$) state
(with excitation energy $E_0$). $S^g$ is the identity matrix and $S^e$
is an arbitrary unitary matrix. $S^e_{11}$ and $S^e_{12}$ correspond
to the reflection and transmission amplitudes respectively.
        \label{fig:fig1}}
\end{figure}

We consider a system at zero temperature with two Fermi surfaces separated by
a barrier with a potential
difference (bias) $V$ applied across the barrier
(see Fig. \ref{fig:fig1}).
The barrier contains a defect,
which exists in one of two states with energy separation
$E_0$. Tunneling through the barrier is assumed to be 
possible only when the defect is in its excited
state. We compute the absorption spectrum
close to the threshold at $\omega_0=E_0-\mbox{Re}(\Delta(V))$,
for frequencies $(\omega-\omega_0) \ll \xi_0$,
where $\xi_0$ is of order the bandwidth and Re$(\Delta(V))$
is real part of the combined energy shift of the two Fermi seas when
the defect is in its excited state. ($\Delta(V)$ is
complex for non-zero $V$ on account of the dissipation
in the system.)

\begin{figure}
{\centering  
\input{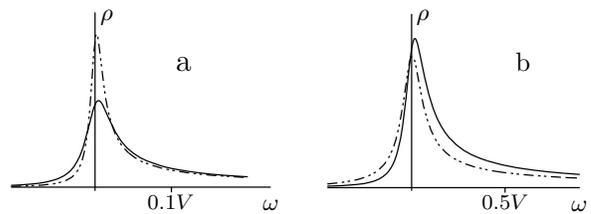} }
\caption{Typical absorption spectra, $\rho(\omega)$, in arbitrary
units for the out-of-equilibrium device sketched in Fig \ref{fig:fig1}.
The spectra depend on $S^e_{11}=\sqrt{R}e^{i\alpha_1}$ and 
$S^e_{22}=\sqrt{R}e^{i\alpha_2}$ (see \ref{eq:rho(omega)} and
\ref{eq:beta'}), where $R$ is the reflection
probability.
(a) $(\alpha_1,\alpha_2)=(0,1.5)$
with $R=0.9$ (broken line) and $R=0.8$ (solid
line). As the reflection probability decreases the spectrum
broadens but retains its asymmetric shape. (b) $R=0.5$ with  $(\alpha_1,\alpha_2)=(0,1.5)$ 
(solid curve) and $(\alpha_1,\alpha_2)=(1.5,0)$ (broken curve).  
The spectrum is sensitive to the difference
in $\alpha_1$ and $\alpha_2$. The difference between the two curves
would show up as differences in the spectra on reversing the bias. }
\label{fig:fig2}
\end{figure}

Using an approach based on that of Nozi\`eres and de Dominicis (ND)
\cite{ND69}, we solve exactly for the asymptotic behavior of the
absorption spectrum in two limiting cases: 
$(\omega-\omega_0)\gg V$ and $(\omega-\omega_0) \ll V$.
Our results  have a simple interpretation in terms
of generalized (complex) phase-shifts at the Fermi energy. Typical
lineshapes for the case $(\omega-\omega_0)\gg V$ illustrating
the dependence on the reflection amplitudes and phases are shown 
in Figure \ref{fig:fig2}.

Our treatment of the problem is based on that
of Muzykantskii and Adamov (MA) for the statistics of charge
transfer in quantum pumps, which uses the relation between
the many-particle response to the changing one-body potential
and the solution of an associated matrix Riemann-Hilbert (RH)
problem \cite{MA03}. This problem was also addressed perturbatively 
and using the ND approach in \cite{CR00,Ng96},
although the results in \cite{CR00} led the authors
to question the validity of the ND approach of \cite{Ng96}
(see also \cite{BB02}). Our 
solution shows  clearly that the ND approach is valid, with the 
earlier difficulties probably associated with an incomplete 
analysis of the matrix RH problem associated with their singular
integral equation.

We characterize the scattering at the
interface between the two subsystems
via the unitary
$2\times2$  matrix, $S(\epsilon,t)$, connecting scattering states in the
two wires for particles with energy $\epsilon$. 
This takes one of two values $S^g$,  and 
$S^e$ depending on whether the the defect is in its
ground ($g$) or excited state ($e$). In the following, 
we will take a row/column 
index equal to one (two) for the left (right) electrode so that
the diagonal (off-diagonal) elements correspond to reflection 
(transmission) amplitudes (see Figure \ref{fig:fig1}).  
We choose the scattering states to be the eigen states of 
the system when the defect is in its
ground state and the barrier is totally reflecting hence:
 $S^g_{ij}=\delta_{ij}$. $S^e$ 
is an arbitrary unitary matrix with
reflection probability $R=|S^e_{11}|^2 < 1$.
We will assume that a negative potential $-V$ ($V>0$)  
has been applied to the left electrode
with respect to the right electrode.

The spectral function, $\rho(\omega,V)$, for absorption by the local
level is given by \cite{ND69}:
\begin{eqnarray}
\rho(\omega,V) & \sim & \text{Re} 
\int_{-\infty}^\infty  \chi (t_f,V) e^{i\omega t_f} dt_f 
\label{eq:rho_definition} \\
\chi (t_f,V) &  = & \langle0|U(t_f,0)|0\rangle.
\label{eq:chi_definition}
\end{eqnarray}
Here $|0\rangle$ is the ground state wavefunction of the complete
system (the filled Fermi seas in the two electrodes
and the defect in its ground state), while
$U(t_f,0)$ is the time-evolution operator
for the system between $t=0$ and $t=t_f$ with the
defect in its excited state.
$\chi(t_f,0)$ is the same as the core hole Green's function computed
in \cite{Mahan67,ND69,CN71}. 

Before discussing the full non-equilibrium
case, we briefly review the known 
equilibrium results. When $V = 0$ 
the response of the system is that of the core hole
problem in a non-separable potential considered in 
\cite{YY82,Matveev-Larkin}
\begin{equation}
\log{\chi(t_f,0)} = -i(E_0-\Delta(0))t_f - \beta \log{it_f\xi_0}
\label{eq:V=0}
\end{equation}
where $\beta= 
\sum_{j=1,2} \left(\frac{\delta_j}{\pi}\right)^2$.
Here  $e^{-i2\delta_j}$ are the
eigen values of $S^e$. 
The threshold is shifted
from $E_0$, the energy separation in the two-level system,
by
$\Delta(0)$, 
which is the shift of the ground state energy of the two Fermi seas 
when the scattering defect is in its excited state.
This standard equilibrium result (\ref{eq:V=0}) 
is well understood in terms
of the low-lying particle-hole excitations created by the rapid
switching of the potential, with the principal contributions to the
logarithm in (\ref{eq:V=0}) from excitations with
frequencies between $t_f^{-1}$ and $\xi_0$. 

When a voltage is applied across the barrier
with the defect in its excited state 
and $R\neq1$, a current will flow and the 
system will become dissipative. 
For $t_f \ll V^{-1}$, the spectral response is
dominated by excitations with frequencies $\omega \gg V$,
involving states which do not sense the potential
drop across the barrier. As a result $\chi(t_f,V)$ 
is unchanged from its value in equilibrium.

When $t_f \gg V^{-1}$, the response 
is controlled by electrons within the band of width
$V$ about the mean Fermi energy. We find that 
\begin{equation} 
\log{\chi(t_f,V)}   =  -i(E_0-\Delta(V)) t_f - \beta' \log{(Vt_f)} + D
\label{eq:Vneq0}  
\end{equation}
Here the function $\Delta(V)$ is given by:
\begin{equation}
\Delta(V)  =  \int_{-\infty}^0  \frac{\mbox{tr}\log{(S(E))}}{2\pi i} dE + 
\int_0^V \frac{\log{(S_{11}(E)) }}{2\pi i} dE  
\label{eq:Delta(V)} 
\end{equation}
This expression (\ref{eq:Delta(V)}) for the (in general complex) 
energy shift of the two Fermi seas, 
when the defect is in its excited state, 
can be thought of as the generalization of Fumi's theorem
\cite{Friedel52,Fumi55} to
the out-of-equilibrium case.
The exponent $\beta'$ in (\ref{eq:Vneq0}) is given by
\begin{equation}
\beta' = \sum_{j=1,2}\left(-\log{(S^e_{jj})}/2\pi i\right)^2 .
\label{eq:beta'}
\end{equation}
The constant term $D$ gives the contribution from excitations
with frequencies between $V$ and $\xi_0$, which do not
sense the potential drop across the barrier. To logarithmic
accuracy \cite{Note_on_cutoff}:
\begin{equation}
D = \beta \log{\xi_0/V}.
\label{eq:D}
\end{equation}

Writing $S^e_{jj}=\sqrt{R}e^{i\alpha_j}$ and comparing the
forms for $\beta$ and $\beta'$
in (\ref{eq:V=0}) and (\ref{eq:beta'}), we see that  the quantity 
$-\log{(S^e_{jj})/2i}=-\alpha_j/2 + i(\log{R})/4\pi$ 
is acting as a complex phase shift. Its real part, $-\alpha_j/2$,
characterizes the scattering in the $j$'th electrode and in (\ref{eq:Vneq0})
describes the effect of particle-hole excitations 
in the band of width $V$ from the Fermi energy. 
Its imaginary part $(\log{R})/4\pi$  relates to the lifetime of the excitation.

The absorption spectrum is found from  the Fourier transform
of $\chi(t_f,V)$ in (\ref{eq:rho_definition}). Measuring
$\omega$ from $\omega_0=E_0-\mbox{Re}(\Delta(V))$, it is given by \cite{note_on_Fourier_transform}:
\begin{equation}
\rho(\omega) \sim \frac{1}{\Omega^{1-\beta'_1}} 
e^{-\beta'_2\phi_\Omega} \sin{\left(\beta'_1\pi- (\beta'_1-1) \phi_\Omega 
-\beta'_2 \log\Omega \right)}.
\label{eq:rho(omega)}
\end{equation}
Here we have defined
$\Omega\exp{i\phi_\Omega} \equiv \omega/V - i(\log{R})/4\pi$
and written
$\beta'=\beta'_1 + i \beta'_2$. 
While the 
dependence on $\beta'_1$ reflects the total overall
scattering on the two sides of the barrier as in equilibrium, 
$\beta'_2$ is proportional to the difference in the phases
of the two reflection amplitudes $S^e_{11}$ and $S^e_{22}$ and
its appearance in (\ref{eq:rho(omega)})
is entirely an out-of-equilibrium effect.

When $R=1$, the term multiplying
$\Omega^{1-\beta'_1}$ in (\ref{eq:rho(omega)}) 
is proportional to the theta function $\theta(\omega)$ and
describes the usual sharp threshold in $\rho(\omega)$. With $R<1$ it
leads to a smearing of the threshold (see Figure \ref{fig:fig2}).
As pointed out in
\cite{CR00}, this broadening of the threshold 
reflects the existence of `negative energy excitations' in the system
involving a hole in the left electrode and a particle in
the right electrode. 
From an experimental point of view, the below threshold
broadening with its functional dependence on the phases
of the reflection amplitudes and its overall
energy scale fixed by the bias are probably the key signatures of
the non-equilibrium effects we are describing.
The sensitivity to the difference in scattering phase shifts
(this difference is proportional to $\beta'_2$) would show up in changes in the 
line shape on reversing the bias and should also be observable.

The derivation of the overlap $\chi(t_f)$ follows quite
closely that of MA  \cite{MA03}.
We introduce the operators  $a_i(\epsilon)$ 
which annihilate particles on the $i$'th side of the barrier
with  energy $\epsilon$ in eigen states of the system
with the defect in its ground state ($S=1$).  
The effect of the time-evolution
operator $U$ acting between $t=0$ and $t_f$
on states $a^\dagger_i|\rangle$, where $|\rangle$ is the true vacuum with no particles,
is given by 
\begin{equation}
Ua^\dagger_i(\epsilon) |\rangle = \sum_{i} \int  d\epsilon'
\sigma_{ij}(\epsilon,\epsilon') a^\dagger_j(\epsilon') |\rangle.
\label{eq:sigma}
\end{equation}
One can show that for states near the Fermi energy
(see \cite{AM01} for example) $\sigma$ is given by:
\begin{equation}
\sigma_{ij}(\epsilon,\epsilon')= e^{-iE_0t_f} \frac{1}{2\pi}
\int_{-\infty}^\infty   S_{ij}(t) e^{i(\epsilon-\epsilon')t} dt
\label{eq:sigma=S}
\end{equation}
provided that the adiabaticity condition
\begin{equation}
\hbar \frac{\partial S}{\partial t} \frac{\partial S}{\partial E} \ll 1
\label{eq:adiabaticity}
\end{equation}
is satisfied. In (\ref{eq:sigma}) $S(t)=S^g$ for 
$t<0$ and $t>t_f$ and $S(t)=S^e$ for $0<t<t_f$ and
we have suppressed the explicit dependence of $S$
on energy. 
When computing the low frequency asymptotics, this
becomes a slow dependence on $(\epsilon + \epsilon')/2$,
and can be neglected.

The overlap $\chi(t_f)$ can be written
\begin{equation}
\chi(t_f) = \langle0|U|0\rangle = \mbox{det}' \sigma
\label{eq:determinant}
\end{equation}
where the prime indicates that the
operator determinant is 
to be taken only over the
occupied states in the two filled Fermi seas.  
This reduces in the equilibrium 
case to the determinant in \cite{CN71}.
With zero chemical potential
in the right electrode and treating the (non-equilibrium) 
Fermi distribution  as the diagonal operator 
$f_{ij}(\epsilon,\epsilon')= \delta_{ij} 
\delta(\epsilon-\epsilon') \theta(-(\epsilon+V(2-i))$ 
allows us to write 
\begin{eqnarray}
\chi(t_f) & = & \mbox{det}(1-f+f\sigma) \label{eq:full_determinant} \\
\log{\chi(t_f)} & = & \mbox{Tr}
\left( \log{(1-f+f\sigma)} - f\log{\sigma} \right) + \mbox{Tr}f\log{\sigma}
\nonumber \\
 & \equiv  &  C(V,t_f) + \mbox{Tr}f\log{\sigma}
\label{eq:log_chi}
\end{eqnarray} 
where the operator determinant is now the full determinant
taken over all states and the trace, Tr, is the trace
over energy and channels. The last term in the expression 
(\ref{eq:log_chi}) can be found by explicitly carrying out the 
integral in (\ref{eq:sigma=S}). This gives
that $\sigma_{ij}(\epsilon,\epsilon') = \delta_{ij}\delta{(\epsilon-\epsilon')}
- X_{ij}(\epsilon-\epsilon')$. The logarithm can then be expanded
as a power series in the matrix $X$
\cite{Note_on_expanding_log_sigma}. After evaluating $X^n$ term by
term and then resumming we obtain:
$\mbox{Tr}f\log{\sigma}=-i(E_0 - \Delta(0))t_f + (V t_f/2\pi i) 
(\log{S})_{11}$. The difference between this and  
$-i(E_0 - \Delta(V))t_f$ in (\ref{eq:Delta(V)}) is contained in the 
function $C(t_f,V)$.

To evaluate $C(V,t_f)$ we introduce
$\widetilde{S}(t,\lambda)$ 
where
\begin{equation}
\widetilde{S}(t,\lambda) =    \exp{(\lambda \log{S(t)}}),
\label{eq:S_matrix}
\end{equation}
so that $\widetilde{S}(t,1)=S(t)$. 
We now apply the following gauge 
transformation: 
\begin{eqnarray}
\mathbf{a}(\epsilon) & \rightarrow & \mathbf{a}(\epsilon,t) =  e^{iLVt} 
\mathbf{a}(\epsilon) 
\label{eq:gauge_transform} \\ 
\widetilde{S}(t,\lambda) & \rightarrow & \widetilde{S}(t,\lambda)=e^{iLVt} 
\widetilde{S}(t,\lambda)
e^{-iLVt} 
\label{eq:S(t,lambda)}
\end{eqnarray}
Here $L$ is the diagonal matrix with $L_{11}=1$ and  $L_{22}=0$. This
has the advantage of eliminating the chemical potential difference
between the two electrodes at the expense of an added time-dependence 
for $\widetilde{S}$ when $t\in [0,t_f]$.
After switching to the time-representation (in which
the trace, Tr,  becomes a trace over channels and an integral
over time) and  substituting
for $\sigma$ from (\ref{eq:sigma=S}), $C(t_f,V)$ can be written
\begin{equation}
C(t_f,V) = \mbox{Tr}
\int_0^\lambda d\lambda  
\left[ \left((1-f+f\widetilde{S})^{-1}f -f\widetilde{S}^{-1}\right)
\frac{d\widetilde{S}}{d\lambda} \right].  
\label{eq:integral_over_lambda}
\end{equation}
Using a parallel argument to that of \cite{MA03}, we find that
\begin{equation}
(1-f+f\widetilde{S})^{-1} = Y_+\left((1-f)Y_+^{-1} + fY_-^{-1}\right).
\label{eq:RH_inverse_matrix}
\end{equation}
where $Y_{\pm}=Y(t\pm i0,\lambda)$.
Here $Y(z,\lambda)$ is  an analytic (matrix) function of complex $z$ 
in the complement of  the cut along the real axis between $z=0$ and 
$z=t_f$, and satisfies: 
\begin{equation}
Y_-Y_+^{-1}  = \widetilde{S}(t,\lambda) \,\,\, \mbox{and} \,\,\,
Y(z,\lambda) \rightarrow  \mbox{const} \,\,\, \mbox{for} \,\,\, |z| \rightarrow
\infty.
\label{eq:RH}
\end{equation}
If there is no tunneling
between electrodes ($S^e$ diagonal),  
this matrix RH problem can be shown to
be the same as the homogeneous part of that solved in \cite{ND69}. 
After substituting 
(\ref{eq:RH_inverse_matrix}) into (\ref{eq:integral_over_lambda}),
using the fact that in the time-representation
(after the gauge transformation
\ref{eq:gauge_transform}) 
$f(t,t')=i(2\pi(t-t'+i0))^{-1}$ and  letting $t'\rightarrow t$ to compute
the trace, Tr, we finally obtain
\begin{equation}
C(t_f,V) = \frac{i}{2\pi} \int_0^1 d\lambda \int_0^{t_f}
\mbox{tr}\left\{\frac{dY_+}{dt} Y_+^{-1} S^{-1}\frac{dS}{d\lambda}\right\} dt.
\label{eq:logY_logS}
\end{equation}
Here tr denotes a trace over channel indices.

Solving for $\chi(t_f,V)$ is equivalent to solving for the 
quantity $Y(z,\lambda)$.   For
small $V$, we can expand the exponential factors
in $\widetilde{S}(z,\lambda)$ (see \ref{eq:S(t,lambda)}) as
$e^{\pm iVz} = 1 \pm iVz$. In this case
\begin{equation}
Y(z,\lambda) = \exp{\left[ \frac{1}{2\pi i}\log{\left(\frac{z}{z-t_f}\right)
\log{\widetilde{S}(z,\lambda)}} \right]}
\label{eq:Y_small_t}
\end{equation}
solves the RH problem. For $|z| \rightarrow \infty$, 
the exponent (and hence $Y$)
tends to a constant as required. If $Vt_f \ll 1$ we can insert this result into 
(\ref{eq:logY_logS}) and compute the integrals over $t$ and $\lambda$.
This yields the equilibrium result (\ref{eq:V=0}).
Although there are corrections
to the equilibrium ($V=0$) solution for  $Y_+$ which are 
linear in $Vt$, these cancel out after
taking the trace in (\ref{eq:logY_logS}). Corrections
to $C(t_f,V)$ can therefore only be of order $(Vt_f)^2$ or higher.

For times $t_f>V^{-1}$, a general solution
to this type of matrix RH problem is not known.
The form (\ref{eq:Y_small_t}) for $Y_+$ is still valid for
$0<t<V^{-1}$ and $t_f>t>t_f-V^{-1}$. The integral over times
close to the branch points of $Y$ then  gives the contribution varying
as $D=\log{(\xi_0/V)}$ in (\ref{eq:D}). However, although
the form for $Y$ in (\ref{eq:Y_small_t})
still satisfies the discontinuity condition along the cut,  the exponent
is unbounded for large $|z|$ and hence (\ref{eq:Y_small_t}) is
useless as a starting point for solving for $Y_+$ for $t\gg V^{-1}$.
Following the derivation of \cite{MA03}, we find that:
\begin{equation}
Y_+(t,\lambda) =  \left[ 
  \begin{array}[c]{rcl}
   \psi_+(t,\lambda) &\hbox{ when }& t< 0 \\
      \begin{pmatrix}
        1 & -\gamma(t,\lambda)  \\ 
        0 & 1
      \end{pmatrix} \psi_+(t,\lambda) & \hbox{  when }& 0 < t < t_f \\
    \psi_+(t,\lambda) &\hbox{ when } &  t_f < t
\end{array} \right.
\label{eq:Y_large_t} 
\end{equation}
is asymptotically correct for $t\gg V^{-1}$.
Here $\gamma(t,\lambda)=\widetilde{S}_{12}(t,\lambda)/
\widetilde{S}_{11}(t,\lambda)$ and $\psi_+(t,\lambda)=\psi(t+i0,\lambda)$ where
\begin{equation}
  \psi(z,\lambda) = \exp \left( \log \frac{z}{z-t_f} 
\left[  \frac{ \log{ \widetilde{S}_{11}/\widetilde{S}^*_{22} } }{4\pi i}\tau_0 
+ \frac{\log{(\widetilde{S}_{11}\widetilde{S}^*_{22})}}{4\pi i}
\tau_3    \right]\right).
  \label{eq:psi}
\end{equation}
The corresponding function $Y(z,\lambda)$  
is not analytic across vertical cuts in the complex $z$-plane
through the points $z=0$ and $z=t_f$,
with discontinuities which decay as $e^{-V|z|}$ or
$e^{-V|z-t_f|}$. (These  factors show that 
we cannot describe the reverse bias case
by taking $V<0$ in (\ref{eq:Y_large_t}). Instead
$Y_+$ takes a different form for negative $V$.)
After inserting the solution (\ref{eq:Y_large_t}) in 
(\ref{eq:logY_logS}) and computing
the integrals over $\lambda$ and $t$, we obtain the first two terms in
(\ref{eq:Vneq0}).
The term obtained after differentiating  $\gamma$ in (\ref{eq:Y_large_t}) 
and adding to the term from $\mbox{Tr}f\log \sigma$ 
in (\ref{eq:log_chi}), leads after some algebra 
to the term $-i(E_0-\Delta(V))t_f$.
Differentiating $\psi_+(t_f,\lambda)$ in (\ref{eq:Y_large_t})    
leads to the  term proportional to $\log{Vt_f}$. 
The constant term is derived using the form (\ref{eq:Y_small_t}) for $Y_+$ 
valid for small $t$ and $t-t_f$ as discussed above.

\bibliographystyle{unsrt}
\bibliography{out_of_eqm}

\begin{thebibliography}{10}

\bibitem{RB94}
D.~C. Ralph and R.~A. Buhrman.
\newblock {\em Phys. Rev. Lett.}, 72:3401, 1994.

\bibitem{KGG01}
M.~Kastner and D.~Goldhaber-Gordon.
\newblock {\em Solid State Communications}, 119:245, 2001.

\bibitem{NCL00}
J.~Nygard, D.H. Cobden, and P.E. Lindelof.
\newblock {\em Nature}, 408:342, 2000.

\bibitem{Mahan67}
G.D. Mahan.
\newblock {\em Phys. Rev.}, 163:1612, 1967.

\bibitem{ND69}
P.~Nozi\`eres and C.T. de~Dominicis.
\newblock {\em Phys. Rev.}, 178:1097, 1969.

\bibitem{CN71}
M.~Combescot and P.~Nozi\`eres.
\newblock {\em Journal de Physique}, 32:913, 1971.

\bibitem{YA70}
G.~Yuval and P.W. Anderson.
\newblock {\em Phys. Rev. B}, 1:1522, 1970.

\bibitem{MA03}
B.~Muzykantskii and Y.~Adamov.
\newblock {\em cond-mat/0301075}, 2003.

\bibitem{CR00}
M.~Combescot and B.~Roulet.
\newblock {\em Phys. Rev. B}, 61:7609, 2000.

\bibitem{Ng96}
T.K. Ng.
\newblock {\em Phys. Rev. B}, 54:5814, 1996.

\bibitem{BB02}
B.~Braunecker.
\newblock {\em cond-mat/0211511}, 2002.

\bibitem{YY82}
K.~Yamada and K.~Yosida.
\newblock {\em Progress of Theoretical Physics}, 68:1504, 1982.

\bibitem{Matveev-Larkin}
K.A. Matveev and A.I. Larkin.
\newblock {\em Phys. Rev. B}, 46:15337, 1992.

\bibitem{Friedel52}
J.~Friedel.
\newblock {\em Phil. Mag.}, 43:153, 1952.

\bibitem{Fumi55}
F.G. Fumi.
\newblock {\em Phil. Mag.}, 46:1007, 1955.

\bibitem{Note_on_cutoff}
There is an additional complex constant of order one not included in
  (\ref{eq:Vneq0}). Its phase (equal to $\pi$ when $V=0$ \cite{ND69}) is fixed
  by the requirement that the absorption spectrum in~(\ref{eq:rho_definition})
  tends to zero for frequencies well below the threshold.

\bibitem{note_on_Fourier_transform}
We should only use the form (\ref{eq:Vneq0}) if the Fourier integral is
  dominated by times $t_f$ with $Vt_f\gg1$. This is the case provided $R\gg
  e^{-4\pi}$.

\bibitem{AM01}
Y.~Adamov and B.~Muzykantskii.
\newblock {\em Phys. Rev. B}, 64:245318, 2001.

\bibitem{Note_on_expanding_log_sigma}
This is essentially the approach used in \cite{CN71} except that all the
  difficulties associated with the cut-off at the Fermi energy have been
  transferred into the term $C(V,t_f)$.

\end{thebibliography}

\end{document}